\documentclass[fleqn,10pt]{wlscirep}
\usepackage[utf8]{inputenc}
\usepackage[T1]{fontenc}
\graphicspath{{figures/}{./}}

\title{Penguin data reanalyzed via Computational Taxonomy}

\author[1]{Thomas B. Michelon}
\author[2]{Avidane C. Caballero}
\author[2,*]{Hsieh Fushing}
\affil[1]{Department of Plant Science, Federal University of Paran\'{a}, Curitiba, PR, Brazil}
\affil[2]{Department of Statistics, University of California at Davis, CA 95616, USA}
\affil[*]{Correspondence: fhsieh@ucdavis.edu}

\keywords{Computational Taxonomy, Scientific Data Analysis, sexual size dimorphism, mate selection, heterogeneity, Pygoscelis}

\begin{abstract}
We employ Computational Taxonomy (CT) to reanalyze the penguin data set \texttt{penguins\_lter} by validating and addressing two biological issues: Sexual Size Dimorphism (SSD) and mate-selection criteria. Via Scientific Data Analysis (SDA) computing, CT constructs a Taxonomic Hierarchy by splitting $Species$ first and then $Sex$, without involving $Island$, to achieve less complexity. This Taxonomic Hierarchy validates SSD as a branch comparison: $(Species, Sex=Male)$-vs-$(Species, Sex=Female)$, upon which SDA explores all potential pieces of associative information from all covariate feature-sets, including interacting effects from order-2 to order-4, and then confirms them via their idiosyncratic reliability checks. The collective of confirmed information pieces are displayed on a heatmap platform to manifest underlying dynamics of SSD with explicit block-structured heterogeneity found within males and females. SSD dynamics is explained through mechanistic dependence pertaining to one chief factor consisting of up to 8 feature-sets: $Body\text{-}Mass$ coupled by combinations of $\{Culmen\text{-}length, Culmen\text{-}depth, Flipper\text{-}length\}$, and two minor factors consisting of low-order combinations of $\{Culmen\text{-}length, Culmen\text{-}depth, Flipper\text{-}length\}$. Such Intra-Sex heterogeneity invalidates all Logistic regression modeling on SSD in the original paper. Further, we explore potential mate-selection criteria through the data-frame of Nest-ID within-species homogeneity.
\end{abstract}

\begin{document}

\flushbottom
\maketitle
\thispagestyle{empty}
\section*{Introduction}
In science and industry, researchers mostly work on complex systems \cite{anderson} contained within our environments or associated with our human societies. The primary avenue of advancing understanding into such systems is indeed analyzing data from such complex systems of interest. In fact, this is the chief path of growing human intelligence regarding our environments and societies.

While a system is typically observable and measurable, however, its underlying dynamics is most often hidden and encoded with complexity \cite{mitchell}. Exploring and then understanding such dynamics becomes the ultimate goal of data analysis in all complex system studies. Recently, we have developed a principled approach called Computational Taxonomy (CT) for analyzing data \cite{CTonIris}. There, CT is illustrated only through the well-known Iris data-set, which is equipped with relatively simple dynamics. For expanding the applicability of CT, we would like to show how CT works on larger systems encoded with more sophisticated complexity in dynamics and embrace wider scopes in biological theories. In this paper, we apply CT to reanalyze the penguin data-set that is derived from a more complex system \cite{gorman} than Iris.

In \cite{gorman}, this penguin data-set was primarily concerned with the ecological and evolutionary issue called Sexual Size Dimorphism (SSD). SSD was discussed through four measurements regarding penguins' body characteristics, together with two measurements of stable isotopes (SI) extracted from blood samples regarding within-species sexual differences through pre-breeding foraging niches. As such, the data set \texttt{penguins\_lter} consists of 6 measurements taken as covariate features, and 3 annotations: $\{Species, Sex, Island\}$, which are three categorical response (Re) features. All SSDs and related biological questions are supposed to be formulated and resolved through the directional associative relationships from covariate (Co) features to response (Re) ones.

SSD is biologically formulated as: What are consistent differences in body characteristics between males and females of the same species? From the perspective of Re-Co dynamics: Sex is the binary response variable, while the 6 measurements are covariate features. From a Statistical perspective, the issue of SSD may be analyzed through logistic regression as done in \cite{gorman}.

However, fundamentally speaking, this simply stated SSD formulation in fact is not well-defined unless the homogeneity assumption can be established within male and female of the same species. In reality, this homogeneity assumption could be violated by $Island$ or any unknown missing response variables. For instance, if males or females of one species are characteristically distinct across three islands: Dream [Adelie and Chinstrap], Torgersen [Adelie], and Biscoe [Adelie and Gentoo] Islands, then SSD has to be resolved with respect to each island.

Particularly, if heterogeneity is revealed within the Adelie species' sex through the Island variable, then it should not be disregarded in analysis of SSD. Such negligence would lead to biased results. So how can we check whether or not heterogeneity is introduced by the Island variable within the Adelie species' sex? This check was not carried out in \cite{gorman}.

What about unknown missing response variables? This is another fundamental issue, or even more so. No one can be sure that no such heterogeneity-capable missing response variables exist, not even biologists. If such missing variables do in fact exist, then the Logistic regression analysis will fail because of the violation of homogeneity assumption required within male and female to sustain the assumed linear structures. Unfortunately, Logistic regression modeling is the primary data analysis tool in \cite{gorman}. As such, this blind-adoption of Logistic regression analysis surely would run into biased or even unscientific results as happened in Chinstrap and Gentoo's SSD analyses reported in \cite{gorman}.

How to systematically check the validity of Adelie's SSD? How to accommodate potential missing heterogeneity-revealing response variables within all three penguin species? We propose CT to resolve such issues systematically and simultaneously.

One fundamental concept of CT is given as follows. In contrast to system complexity brought out in the fashion of heterogeneity-vs-homogeneity mapping on the response side, there is one kind of dynamic complexity manifested in the fashion of mechanistic dependence arising from the covariate (Co) side. Ideally, one homogeneity is specified by one phase of one chief factor within the complex system's underlying dynamics. It is essential to elaborate how a homogeneity within the response side corresponds to a dynamic phase revealed through mechanistic dependence among members of a chief factor on the covariate side. Mechanistic dependence refers to the phenomenon that multiple categories belonging to multiple covariate feature-sets exclusively share the same set of study-subjects.

That is, mechanistic dependence reveals how and where multiple feature-sets work together. That is why mechanistic dependence manifests the system's dynamics from a certain locality-aspect. Further, the exclusiveness of study-subject sharing among multiple mechanistically dependent feature-sets indicates a key aspect of the system's underlying dynamics, which is called a chief factor. Such a chief factor, framed by strong mechanistic dependence across several localities, indeed brings out intrinsic components of the system's underlying dynamics. We can explain the system's dynamics via a chief factor regarding how a group of feature-sets work together, and where they work together. It is noted that a chief factor does not necessarily have a constant, but consistent, membership across all phases.

The commonly used concepts of correlation in Statistics or dependence in probability theory are variable-to-variable relations of global nature. They are not flexible enough to describe system dynamics, which varies from one locality to another locality. We illuminate such a critical difference between CT and Statistical analysis. In resolving species-specific SSD in \cite{gorman}, the primary data analysis is performed by applying Logistic regression framed by one linearity of the four visible covariate feature-measurements. Ideally, this linearity with properly chosen parameter values will give rise to two ``global phases'' matching separately with the binary response variable's two categories: male and female. Thus, under the logistic regression modeling, to facilitate a global binary-matching requires one version of global homogeneity within male and another version of global homogeneity within female.

This kind of homogeneity assumption of global scale is unrealistic. Ignoring this assumption when applying this statistical analysis is an unscientific act. Further, due to this linear structure, all interacting effects are nearly completely left out. That is, such Logistic regression analyses are likely based on wrong modeling structures by leaving out potentially critical relational patterns. In fact, all regression models, including Logistic regression model, always seek for mutual independent variables to form a linear structure. In sharp contrast, the most vital part of CT is a collective of mechanistic dependence patterns, with which various locality-based system dynamics are manifested.

\section*{Materials and methods}
In this paper, we take the Computational Taxonomy approach \cite{CTonIris}. We first construct a taxonomic hierarchy in a data-driven fashion without relying on any assumptions or man-made structures. Such a taxonomic hierarchy progressively maps out heterogeneity-vs-homogeneity via tree geometry. As such the entire data-set is sequentially split and partitioned, and to the end a resultant collection of data subsets called classes are recognized and found. As a homogeneity-entity, one class is one ending node of this hierarchy. With respect to this taxonomic hierarchy, the biological meaning of species-specific SSD would be established as branch-vs-branch comparison, which is then translated into a properly defined Re-Co dynamic.

Within a Re-Co dynamic, Scientific Data Analysis (SDA)~\cite{omotayo} is applied to bring out the associative information based manifestation of SSD. An early version of SDA can be found in \cite{FCC23}. Essentially, after clustering all quantitative feature-sets into a categorical form \cite{FR2018}, SDA utilizes contingency tables to explore associative information through row-wise and/or column-wise localities of covariate feature-sets. That is, each piece of associative information is a conditional distribution of categorical response-variable given a category-locality of a focal feature-set. Since each contingency table fully reveals its randomness, two ensembles of contingency tables are simulated to capture the observed randomness and null randomness. As such, each associative information piece is equipped with a finite-sample precision manifested through two Shannon entropy distributions~\cite{cover,CTonIris}. Their overlapping area is taken as the information piece's idiosyncratic measurement of finite-sample precision and reported along the bar-column within each heatmap throughout this paper. This finite-sample precision is a way of carrying out Kolmogorov's randomness proper concept~\cite{kolmogorov}.

This SDA is also applied for exploring the mate-selection criteria. In this context, our SDA approach is sharply contrasting the classic Chi-squared Statistical approach upon the contingency table platform. This biological question was not considered in the original paper \cite{gorman}.

The original paper provided the following four feature measurements via four observable and measurable body characteristics as:\\
$\{culmen\text{-}length (CL), culmen\text{-}depth (CD), flipper\text{-}length (FL), body\text{-}mass (BM)\}$, and two stable isotopes from blood samples: $\{\delta^{15}\mathrm{N}, \delta^{13}\mathrm{C}\}$, are also collected from each penguin-pair, when they produced one egg in their nest. These two isotopes' measurements are collected to address the potential within-species differences between male and female regarding their intakes during the reproducing stage. They are taken as two covariate features for a dynamics distinct from SSD. Therefore, SSD is referred to the four body character-based covariate features, while within-species sexual differences on foraging is referred to the two stable isotopes as covariate features.

\section*{Computational Taxonomy}
In Computational Taxonomy, Taxonomic Hierarchy embedded within this Penguin data is constructed via a practical computing protocol given as follows. Under the categorical response setting of $\{Species, Sex, Island\}$, for the sake of achieving the ideal collection of classes with least complexity, we begin by checking which one of the three response features alone can achieve a partition on the entire data set that is closest to the ideal one. The best one is the one that its categories are most evidently separable with respect to pieces of associative information provided by the 4 body-characteristic covariate features. That is, one response feature individually gives rise to one Re-Co dynamic.

Each Re-Co dynamic is explored fully by Scientific Data Analysis (SDA) for its full spectrum of pieces of relational associative information, including order-1 major effects of the 4 covariate features and their interacting effects from order-2 to order-4. All computed and confirmed associative information pieces are collected and displayed in a heatmap platform. As a bipartite network of binary participation of penguins, a heatmap has all study-subjects being arranged along its column-axis and all selected associative information pieces being arranged along its row-axis.

Along our computing protocol, a series of heatmaps would clearly reveal a progressive series of heterogeneity-vs-homogeneity maps partitioning the entire data set. Each heterogeneity-vs-homogeneity map becomes evident when such a heatmap is sustained by block-structures framed by two hierarchical clustering trees built on row- and column-axes.

\begin{figure}[h!]
 \centering
 \includegraphics[width=1.0\textwidth]{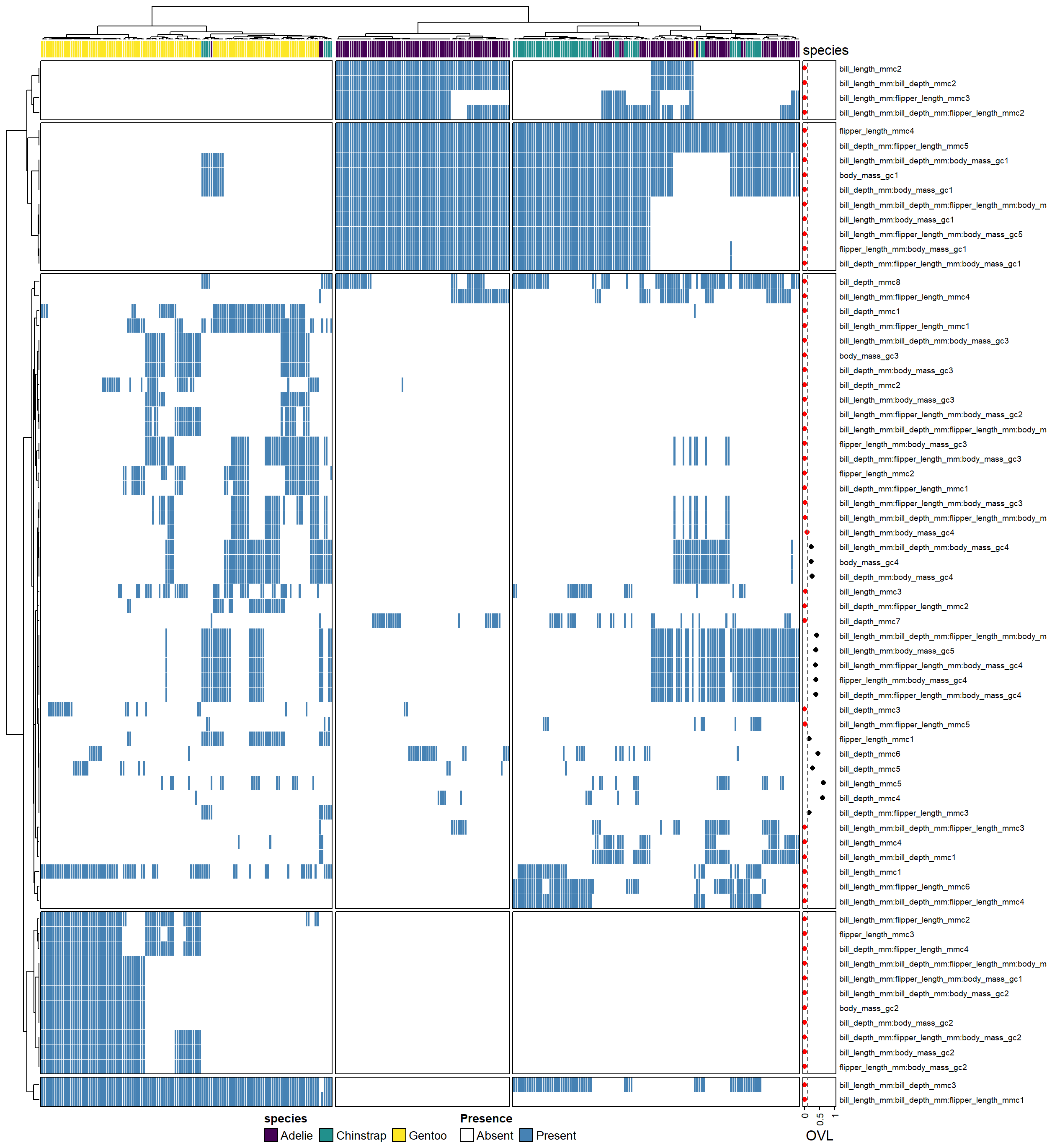}
 \caption{Heatmap of Species as a response variable. Each associative information piece's individual overlapping areas (red-dot for being less than 0.1, blue-dot for being larger than 0.1) reported along the bar-column on the right hand side of heatmap.}
 \label{Species}
 \end{figure}

\begin{figure}[h!]
 \centering
 \includegraphics[width=1.0\textwidth]{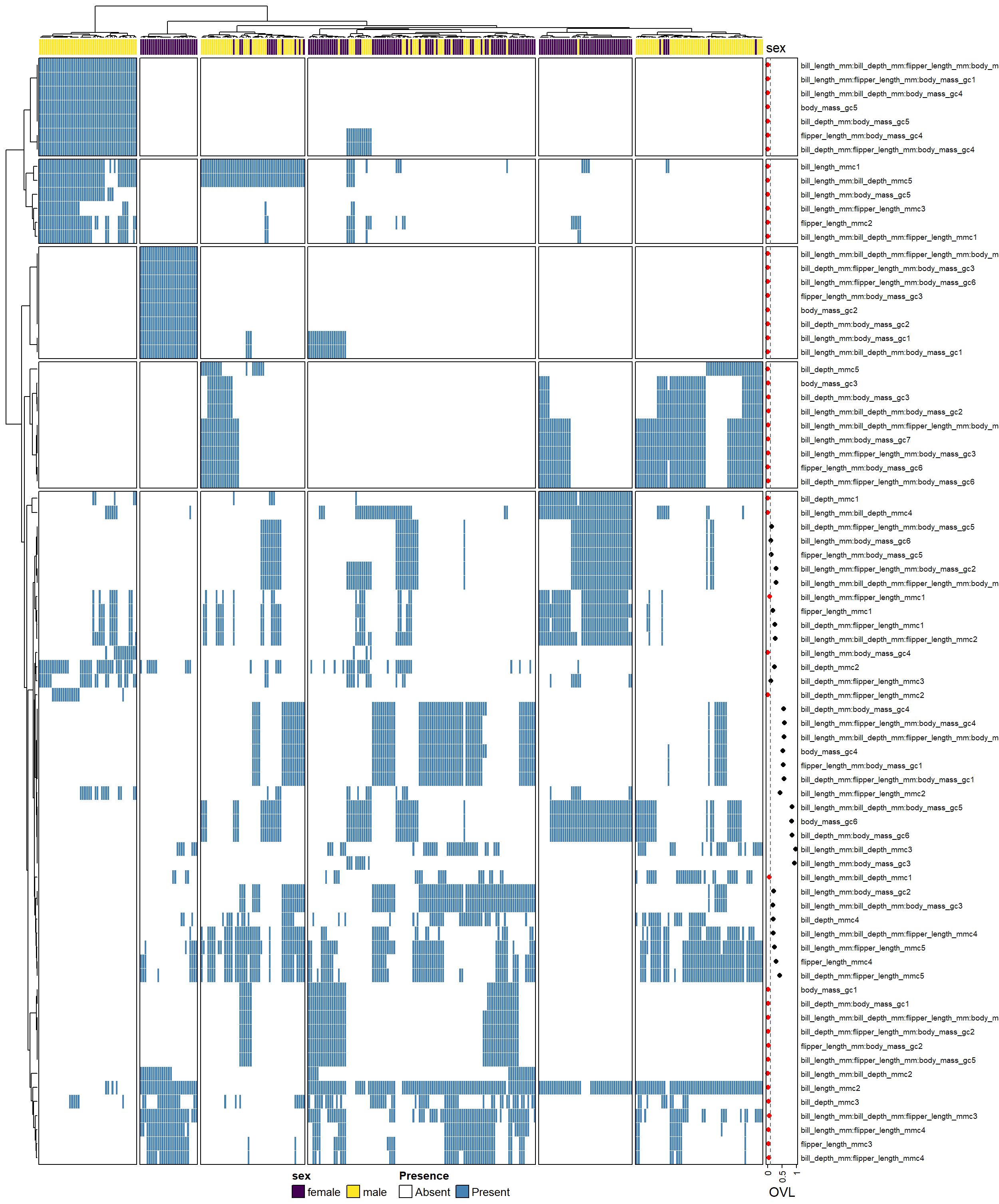}
 \caption{Heatmap of Sex as a binary response variable. }
 \label{Sex}
 \end{figure}

\begin{figure}[h!]
 \centering
 \includegraphics[width=1.0\textwidth]{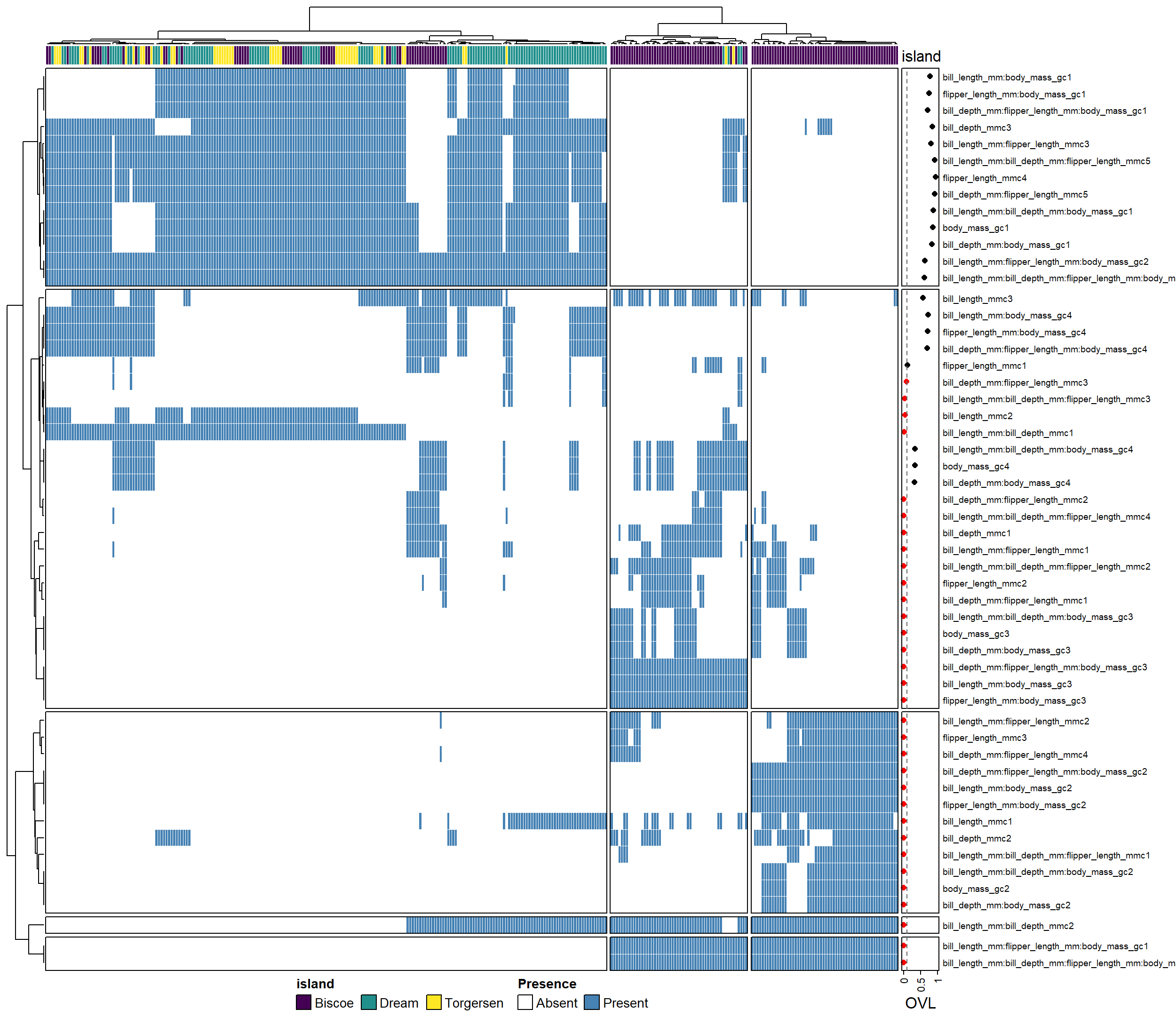}
 \caption{Heatmap of Island as a triplet response variable. }
 \label{Island}
 \end{figure}

\paragraph{The 1st level of Taxonomic Hierarchy:}
Taking each of the three individual response features as one response variable and respectively applying SDA, we construct three heatmaps, respectively. The heatmap shown in Fig.~\ref{Species} is for $Species$, the heatmap in Fig.~\ref{Sex} for $Sex$ as binary response variable, and the heatmap in Fig.~\ref{Island} for $Island$ as a triplet response variable.

\begin{figure}[h!]
 \centering
 \includegraphics[width=1.0\textwidth]{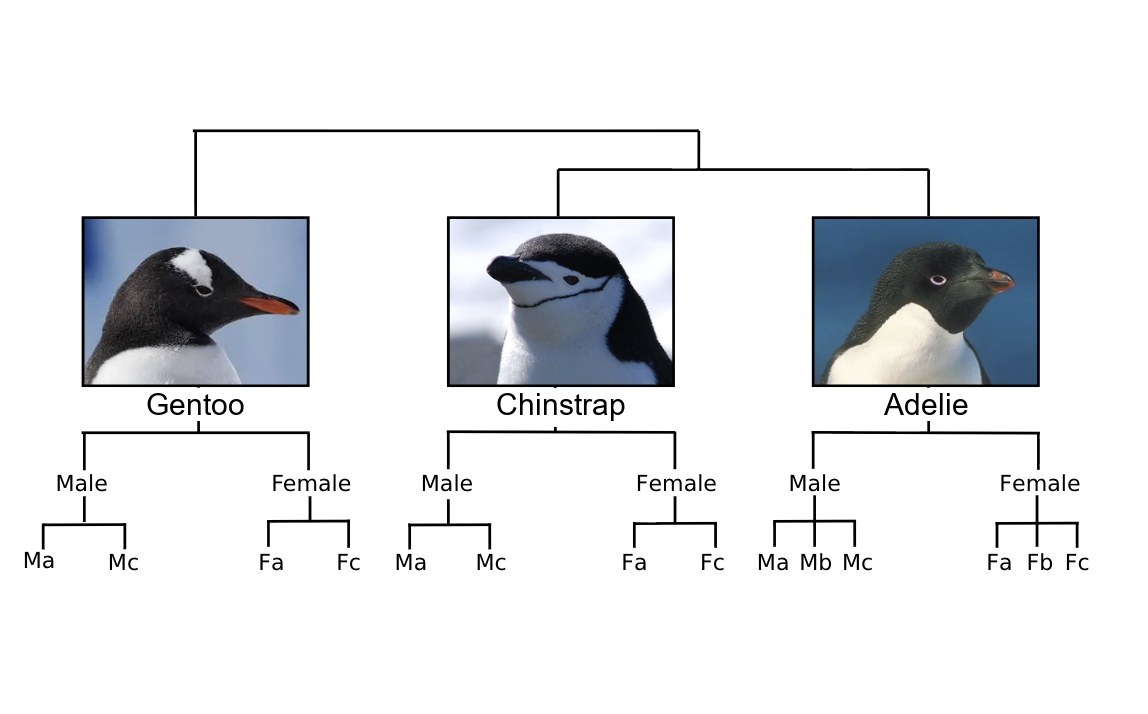}
 \caption{Expanded taxonomic hierarchy on penguin data. All species-specific classes encoded with IDs: $\{Ma, Mb, Mc, Fa, Fb, Fc\}$, are indicated on the Hierarchical Clustering tree-branches in Fig.~\ref{SSDAdelie}, Fig.~\ref{SSDGentoo} and Fig.~\ref{SSDChinstrap}.}
 \label{hierarchy}
 \end{figure}

Then, by comparing these three SDA resultant heatmaps, we can explicitly see and clearly explain which response feature achieves the most concise heterogeneity-vs-homogeneity map. Species is chosen as the winner with all its three categories being well separated, as shown in Fig.~\ref{Species}. And we also recognize that $Adelie$ and $Chinstrap$ are closer to each other than to $Gentoo$. A tree having three Species-branches is shown via the first tree-level in Fig.~\ref{hierarchy}.

\begin{figure}[h!]
 \centering
 \includegraphics[width=1.0\textwidth]{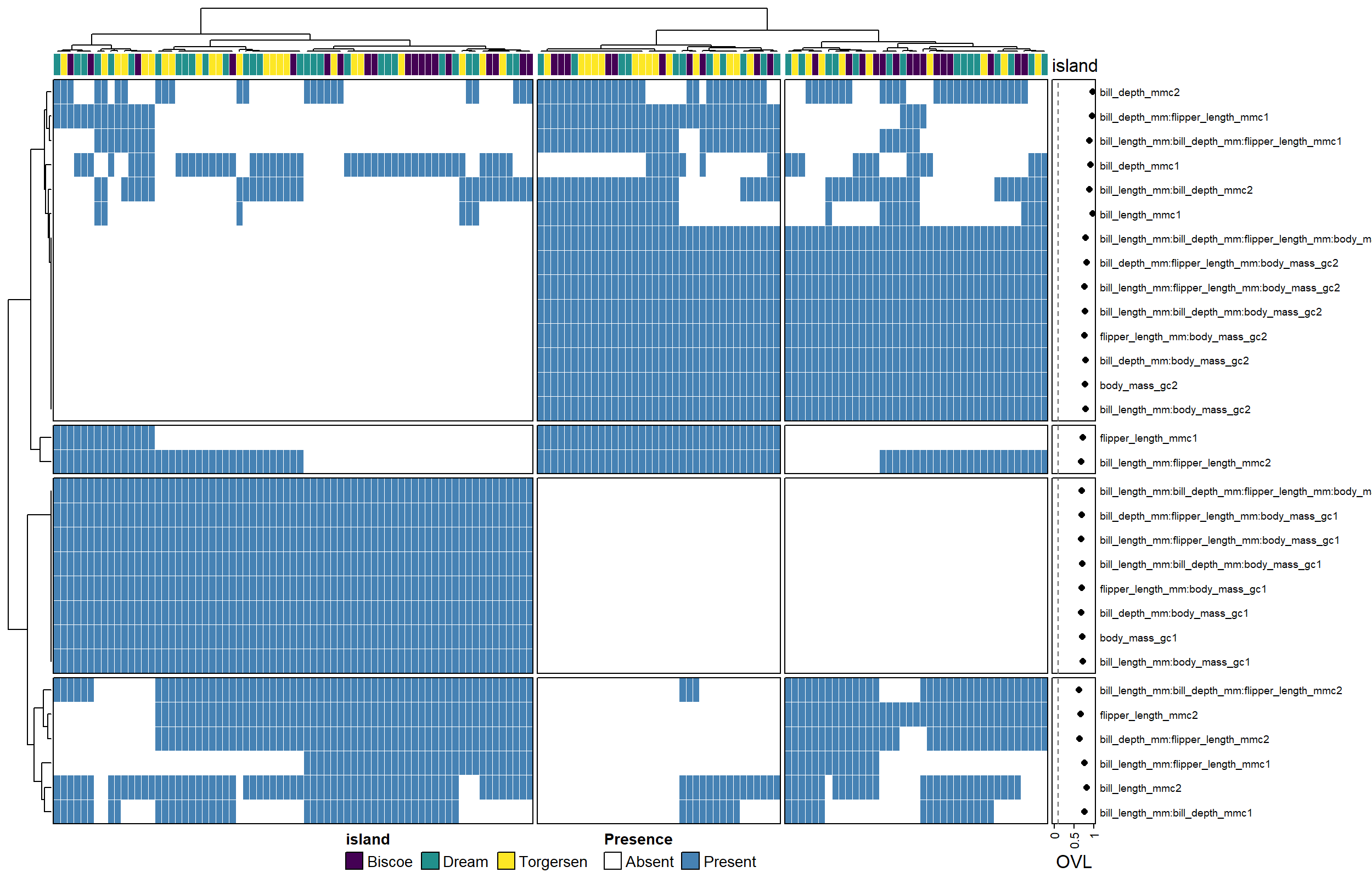}
 \caption{Heatmap of $Adelie$ being split via three categories of $Island$. }
 \label{Adelieisland}
 \end{figure}

\begin{figure}[h!]
 \centering
 \includegraphics[width=1.0\textwidth]{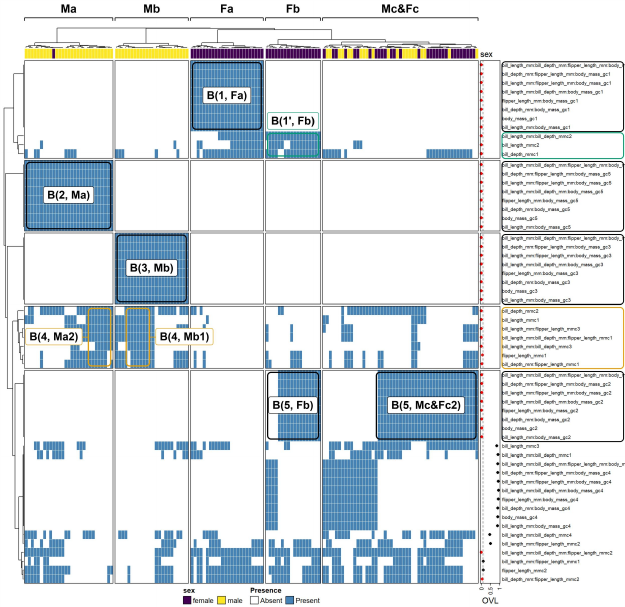}
 \caption{SSD of Adelie penguins. }
 \label{SSDAdelie}
 \end{figure}

\paragraph{The 2nd level of Taxonomic Hierarchy.}
We then go on to explore the potential splitting of each of the three species-branches individually with respect to $\{Sex, Island\}$ via the SDA computations again. Since Adelie is the only species found on three islands, heatmap comparisons pertaining to $\{Sex, Island\}$ are performed on the Adelie species: the heatmap for $Island$ in Fig.~\ref{Adelieisland} against the heatmap of $Sex$ in Fig.~\ref{SSDAdelie}.

\begin{figure}[h!]
 \centering
 \includegraphics[width=1.0\textwidth]{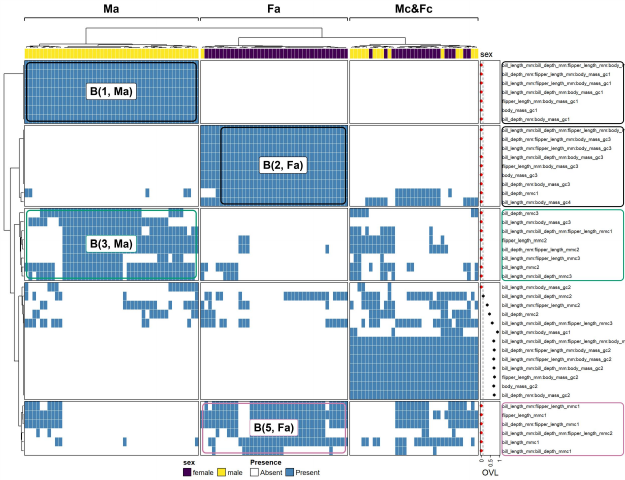}
 \caption{SSD of Gentoo penguins. }
 \label{SSDGentoo}
 \end{figure}

\begin{figure}[h!]
 \centering
 \includegraphics[width=1.0\textwidth]{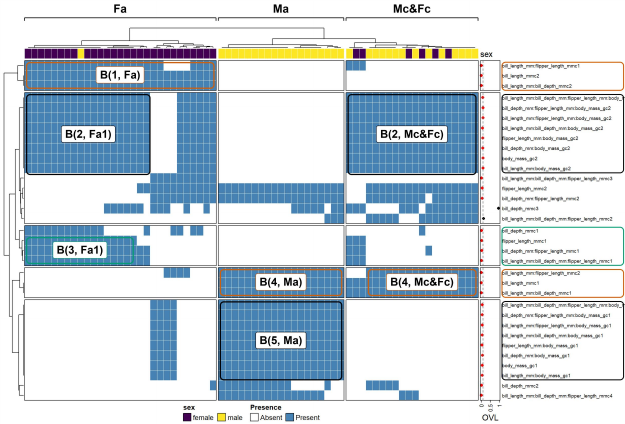}
 \caption{SSD of Chinstrap penguins. }
 \label{SSDChinstrap}
 \end{figure}

Based on heatmap in Fig.~\ref{SSDAdelie}, it is clearly seen that Species Adelie should be further split with respect to $Sex$ to be closer to the ideal heterogeneity-vs-homogeneity map. Likewise, heatmap in Fig.~\ref{SSDGentoo} clearly indicates that Species Gentoo needs to be further split by $Sex$. And heatmap in Fig.~\ref{SSDChinstrap} indicates the Species Chinstrap needs to be split further with respect to $Sex$.

In summary, $Sex$ is selected to further split each species: $\{Adelie, Chinstrap, Gentoo\}$ individually. Up to this point, the taxonomic hierarchy has been constructed as a tree with 3 species-branches, each of which is then split into two with respect to Male and Female. So far, the computed taxonomic hierarchy has grown with 6 species-sex branches shown via first and second tree-levels of the tree-geometry in Fig.~\ref{hierarchy}.

\paragraph{The 3rd level of Taxonomic Hierarchy.}
In this penguin data set, there are 10 observed response-categories, not $18 (=3 \times 3 \times 2)$. Gentoo and Chinstrap are each observed on a single island. Again only Adelie is observed across three islands. The last step of constructing Taxonomic Hierarchy is to determine whether the Adelie's Male and Female categories require further splitting with respect to the three islands, or are kept in their current states.

\begin{figure}[h!]
 \centering
 \includegraphics[width=1.0\textwidth]{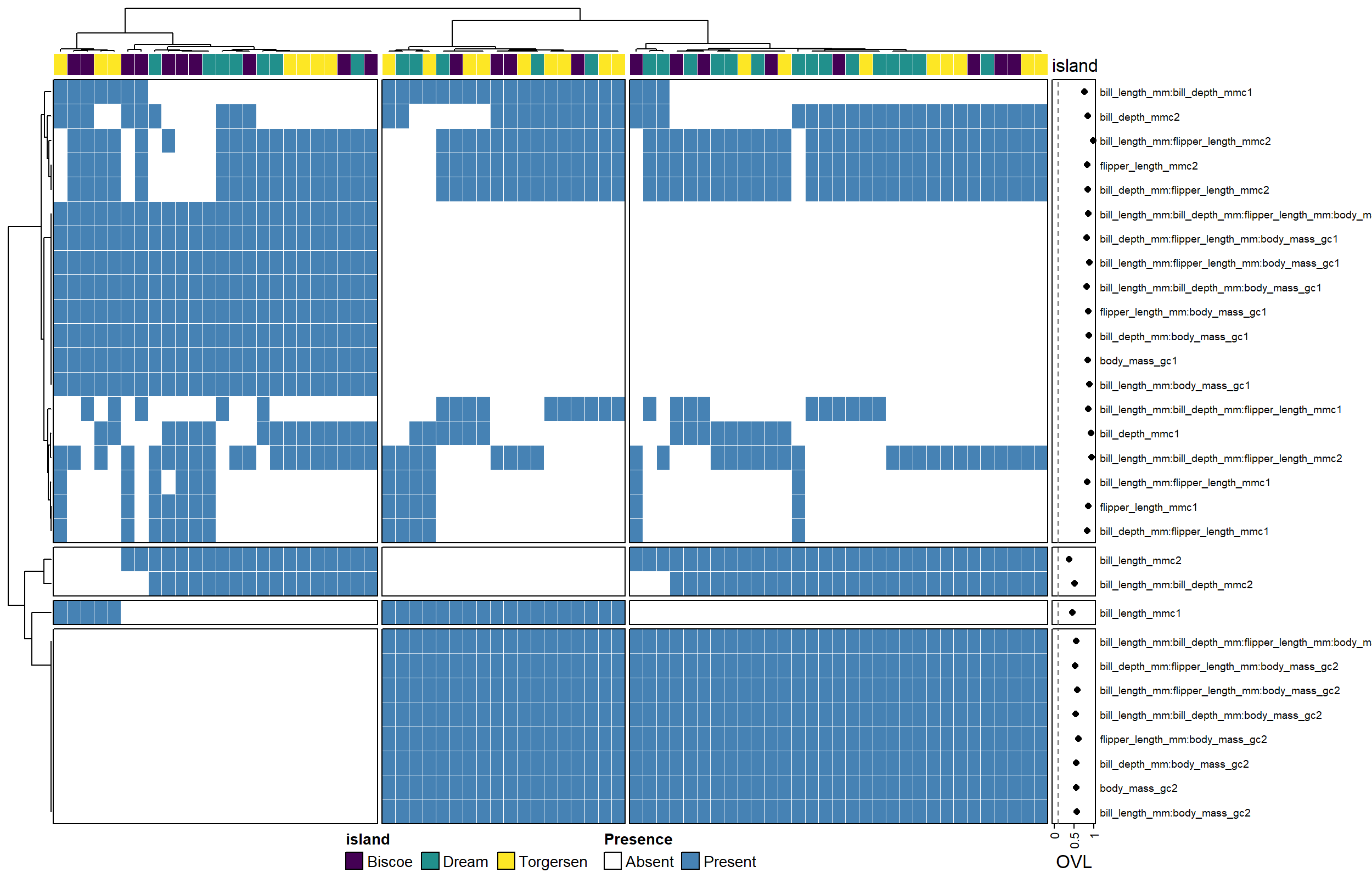}
 \caption{Heatmap of female Adelie w.r.t Island. }
 \label{Adeliefemaleisland}
 \end{figure}

\begin{figure}[h!]
 \centering
 \includegraphics[width=1.0\textwidth]{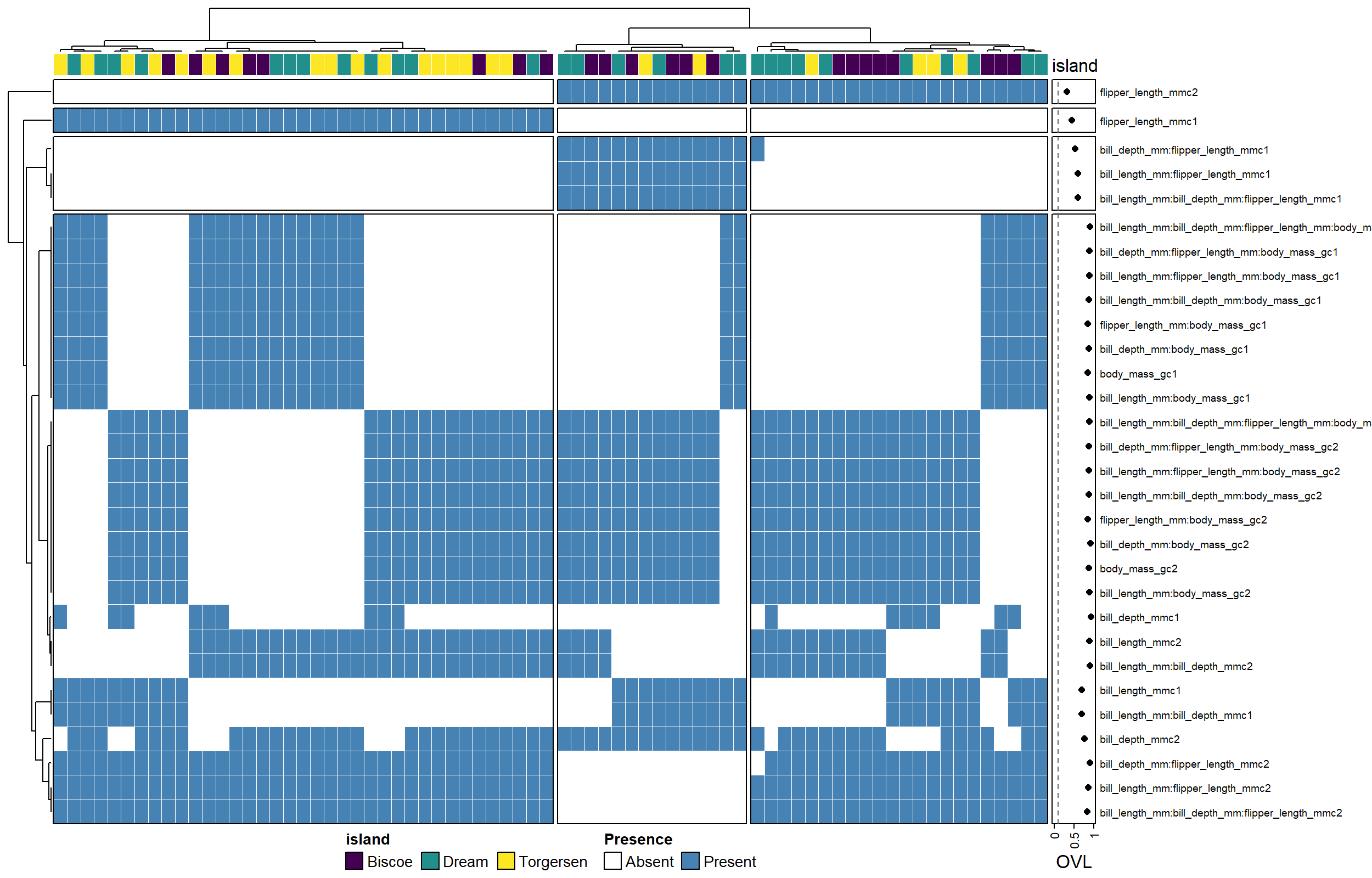}
 \caption{Heatmap of male Adelie w.r.t Island.}
 \label{Adeliemaleisland}
 \end{figure}

According to SDA computations reported in two heatmaps of Fig.~\ref{Adeliefemaleisland} and Fig.~\ref{Adeliemaleisland}, neither Adelie-Male nor Adelie-Female need to be further split. As such, the Taxonomic Hierarchy remains a binary tree having 6 Species-Sex branches.

In summary, the construction of taxonomic hierarchy is necessary for justifying the biological content of SSD as being defined purely species-specific. Our resultant taxonomic hierarchy indeed excludes $Island$ from its structure. Hypothetically,
if the resultant taxonomic hierarchy was shown indeed to embrace $Island$, then SSD would have been defined rather differently. We have resolved the first task in this reanalysis on penguin data.

\section*{SSD across three species}
After building the Taxonomic Hierarchy, each species-specific SSD is biologically validated and translated into a branch-comparison. Though the ending node $(Species, Sex)$ of this Taxonomic Hierarchy would not be further split with respect to $Island$ to gain higher degrees of homogeneity, it doesn't mean that each $(Species, Sex)$ category is indeed homogeneous with respect to the four covariate features. In this section, we demonstrate this computed fact: Each $(Species, Sex)$ branch of taxonomic hierarchy indeed embraces varying degrees of heterogeneity.

Unfortunately, the statistical analysis via logistic regression modeling reported in \cite{gorman} assumed and relied on this homogeneity assumption on every $(Species, Sex)$ category. This assumption indeed is violated across all 6 $(Species, Sex)$ categories. Hence, the reported results and conclusions there are either incomplete or questionable. For instance, the authors reported that no interacting effects could be found. This statement will be shown to be false below. Indeed, interacting effects of order-2 to order-4 are evidently seen across all Species' SSD. The underlying reason is that the assumed linearity structure in Logistic regression tends to pick up independent covariate variables via the maximum likelihood approach. As such, Logistic Regression modeling is not capable of accommodating interacting effects of varying orders coherently.

In sharp contrast, our computational results, their interpretations, and explanations via Computational Taxonomy manifest the critical importance of ``mechanistic dependence'' for understanding the dynamics underlying all species' SSD. This fundamental concept of mechanistic dependence is the chief source of essential information contents in penguin data regarding the three Species-specific SSD through three heatmaps, respectively. The heatmap for Adelie-male-vs-Adelie-female is shown in Fig.~\ref{SSDAdelie}, for Gentoo-male-vs-Gentoo-female in Fig.~\ref{SSDGentoo} and Chinstrap-male-vs-Chinstrap-female shown in Fig.~\ref{SSDChinstrap}.  These heatmaps are taken as exact answers to Species-specific SSD.

Such explicitly visible answers are to be interpreted through the block-chain-patterns as shown in all three block-sustained structures. Each block of 1's or 0's is framed by a cluster of penguins on the column-axis and one cluster of confirmed associative information pieces of various orders on the row-axis. Biologically, each block of 1's manifests mechanistic dependence among covariate feature-sets at their localities. The confirmed associative information is revealed through this mechanistic dependence by sharing the same cluster of penguins. This block conveys very critical knowledge on the underlying dynamic of the species-specific SSD. While a block of 0's also manifests mechanistic dependence through absence with respect to another cluster of associative information. Therefore, a cluster of penguins would explicitly be characterized by a vertical block-chain. Overall, such a vertical block-chain characterization is multiscale.

Such heatmaps reveal information about which branches need further splitting to have homogeneity required by a class. That is, a class is meant to be uniquely characterized by one block-chain. Any branch consisting of multiple individual characterizations via multiple block-chains indeed uncover its within-branch heterogeneity-homogeneity maps. In other words, from the Computational Taxonomy perspective, a class is an entity that can't be further split into multiple classes. Based on such block-chain based mechanistic dependence, we show our computational results for SSD being fundamentally distinct from the original paper~\cite{gorman}.

We first discuss the SSD of Adelie via heatmap in Fig.~\ref{SSDAdelie}. This Adelie-specific SSD embraces the highest complexity of block-chain based mechanistic dependence via the Hierarchical Clustering (HC) tree on the column-axis of the heatmap. Five vertical block-chains stand out most evidently, corresponding to five marked penguin cluster-branches of Adelie: $\{Ma, Mb, Fa, Fb, Mc\&Fc\}$. Here $Mc\&Fc$ denotes a branch consisting of a mixture of males and females, whose two sub-branches are denoted as $Mc\&Fc1$ and $Mc\&Fc2$, respectively. Each one of these five vertical block-chains consists of one block of 1's:\\ $\{B(2, Ma), B(3, Mb), B(1, Fa), B(5, Fb), B(5, Mc\&Fc2)\}$. And all these blocks of 1's are marked with the same black circle-around. This common black-colored circle-around is meant to indicate that these five blocks of 1's are commonly defined by 8 $(=2^3)$ feature-sets: all possible combinations of covariate features of $\{CD, CL, FL\}$ coupling with covariate feature $BM$, which is called the chief factor.

This chief factor is computationally identified as the most critical underlying dynamics of SSD, as would be seen throughout the other two species' heatmaps. In the heatmap in Fig.~\ref{SSDAdelie}, the above 5 blocks of 1's are characterized by four different phases of mechanistic dependence pertaining to the chief factor. The last two $\{B(5, Fb), B(5, Mc\&Fc2)\}$ share the same phase of the chief factor.

The heatmap in Fig.~\ref{SSDAdelie} also consists of three blocks with a non-black colored circle:\\ $\{B(4, Ma2), B(4, Mb1), B(1', Fb)\}$. Here, a non-black color encodes a minor factor, which is a collection of feature-sets. That is, both $\{B(4, Ma2), B(4, Mb1)\}$ (in yellow) are defined by the same phase of a minor factor, while $B(1', Fb)$ (in green) is defined by one phase of another minor factor.

These two minor factors also play a very fundamental role in discovering the heterogeneity-vs-homogeneity map pertaining to the Adelie's SSD. The presence of block $B(4, Ma2)$ of 1's strongly indicates that $Ma$ should be further split into two. Likewise, the presence of $B(4, Mb1)$ indicates further splitting into two for $Mb$. Interestingly and importantly, the presence of $B(1', Fb)$ is only coupled with $B(5, Fb)$, but not $B(5, Mc\&Fc2)$. Therefore, $Fb$ is characterized differently from $B(5, Mc\&Fc2)$, despite sharing the same phase of the chief factor. More concisely, $Mc\&Fc$ should be split into $Mc\&Fc1$, a block of 0's, and $Mc\&Fc2$, a block of 1's.

As such, it is an extremely amazing and interesting fact that we discover the underlying dynamics of Adelie's SSD through the block-sustained heatmap in Fig.~\ref{SSDAdelie}. Within this heatmap, the chief factor and two minor factors collectively characterize the heterogeneity-vs-homogeneity map for males and females of the Adelie species through mechanistic dependence across varying phases. It is worth mentioning that $BM$ plays a dominant role within the chief factor, as does the underlying dynamics of Adelie's SSD.

Next, we discuss the underlying dynamics of Gentoo's SSD based on heatmap of Fig.~\ref{SSDGentoo}. This SSD dynamic is much simpler than that of Adelie. It consists of two blocks of 1's: $\{B(1,Ma), B(2, Fa)\}$, which are characterized by two distinct phases of the same chief factor found in Adelie's SSD. That is why they are marked with black-colored circle. It is noted that this chief factor consists of 7, not 8, covariate feature-sets by missing $(BM, CL)$.

In sharp contrast, the cluster $Mc\&Fc$ on the HC-tree is characterized by the absences of the phases of the chief factor. Further, the two blocks $\{B(1,Ma), B(2, Fa)\}$ are coupled with two blocks: $\{B(3, Ma), B(5, Fa)\}$ pertaining to two distinct minor factors, respectively, in characterizing $Ma$ and $Fa$.

Finally, we discuss the underlying dynamics of Chinstrap's SSD based on the heatmap in Fig.~\ref{SSDChinstrap}. The heatmap sustains 7 marked blocks of 1's with various colored circle-arounds. That is, Chinstrap's SSD dynamics is much more complex than Gentoo's. It is less complex than Adelie's, but its two minor factors play important roles in characterizing $\{Ma, Fa, Mc\&Fc\}$.

Among the 7 blocks, 3 blocks: $\{B(2, Fa1), B(2, Mc\&Fc), B(5, Ma)\}$ are commonly marked with black, meaning they are characterized by the three distinct phases of the chief factor. It is essential to note that block $B(2, Fa1)$ is coupled with block $B(3, Fa1)$ (in green). Together, this coupling strongly indicates that $Fa$, which is characterized by block $B(1, Fa)$ (in red), should be split into $Fa1$ and $Fa2$. Though blocks $B(2, Fa1)$ and $B(2, Mc\&Fc)$ share the same phase of the chief factor, the latter is also characterized by block $B(4, Mc\&Fc)$ of 1's (also in red). That is, the two blocks: $B(1, Fa)$ and $B(4, Mc\&Fc)$ are subject to two distinct phases of the same minor factor. As for the block $B(5, Ma)$, it is also characterized by the same minor factor (in red).

After describing the dynamics of SSD across three penguin species, it is necessary to compare the results and conclusions in the original paper. The following two statements are stated in the Results and Conclusion, respectively:
\begin{description}
\item[Results:]``However, Chinstrap penguins were more sexually dimorphic in culmen and flipper features than
Adelie and gentoo penguins. Adelies and Gentoos were more sexually dimorphic in body mass than chinstraps.''
\item[Conclusion:]``Chinstraps were most sexually size dimorphic followed by Gentoos and Adelies.''
\end{description}
These two statements are almost completely incoherent with the block-chain patterns observed through the three heatmaps summarizing SSDs of the three species by nearly completely missing the chief and minor factors. To a certain extent, the Chinstrap's SSD dynamics apparently are enhanced by order-2 to order-4 mechanistic dependence pertaining to two minor factors. However, the cluster $Ma$ of males and the $Fa1$ of females are strictly characterized by two distinct phases of mechanistic dependence of the chief factor. This is one major discovery. As for the two minor factors involved in Chinstrap's SSD dynamics, it is critical to explicitly define which kinds of mechanistic dependence are involved in defining these two minor factors and which phases characterize the clusters of Chinstrap's males and females.

To a great extent, the computed chief factor is an important discovery. Its distinct phases will define distinct clusters and sub-clusters of male and female penguins across the three species. Its effects in defining SSD dynamics are evident and critical, especially in Adelie and Chinstrap. Equally important is that these three penguin species' SSD dynamics can be explained in detail by combining the effects via mechanistic dependence of various orders pertaining to the chief factors and various minor factors.

It is essential to reiterate that the most evident patterns of SSD are seen through the block-based complexity in heatmaps driven by the mechanistic dependence pertaining to the chief factor and minor factors. Adelie and Chinstrap reveal the most complex heterogeneity-vs-homogeneity map in their heatmaps. Among the three species, the Gentoo has the least complex one. Though this chief factor is driven by the dominance of $BM$, the collection of block-based mechanistic dependence patterns is clear cut across all three heatmaps. That is, the primary driving force in SSD pertains to this chief factor. The minor factors also play an important supporting role.

Finally, upon the aforementioned heterogeneity within categories $(Species, Sex)$, the full version of Taxonomic Hierarchy is accordingly expanded and constructed according to the heterogeneity identified through SSD-heatmaps as shown in Fig.~\ref{hierarchy}. This version of Taxonomic Hierarchy provides a heterogeneity-homogeneity map. Each ending tree-node is an identified class. Upon this collection of classes and their tree-geometry, all scientific questions would be translated into branch-vs-branch comparisons pertaining to this Taxonomic Hierarchy.

\begin{figure}[h!]
 \centering
 \includegraphics[width=1.0\textwidth]{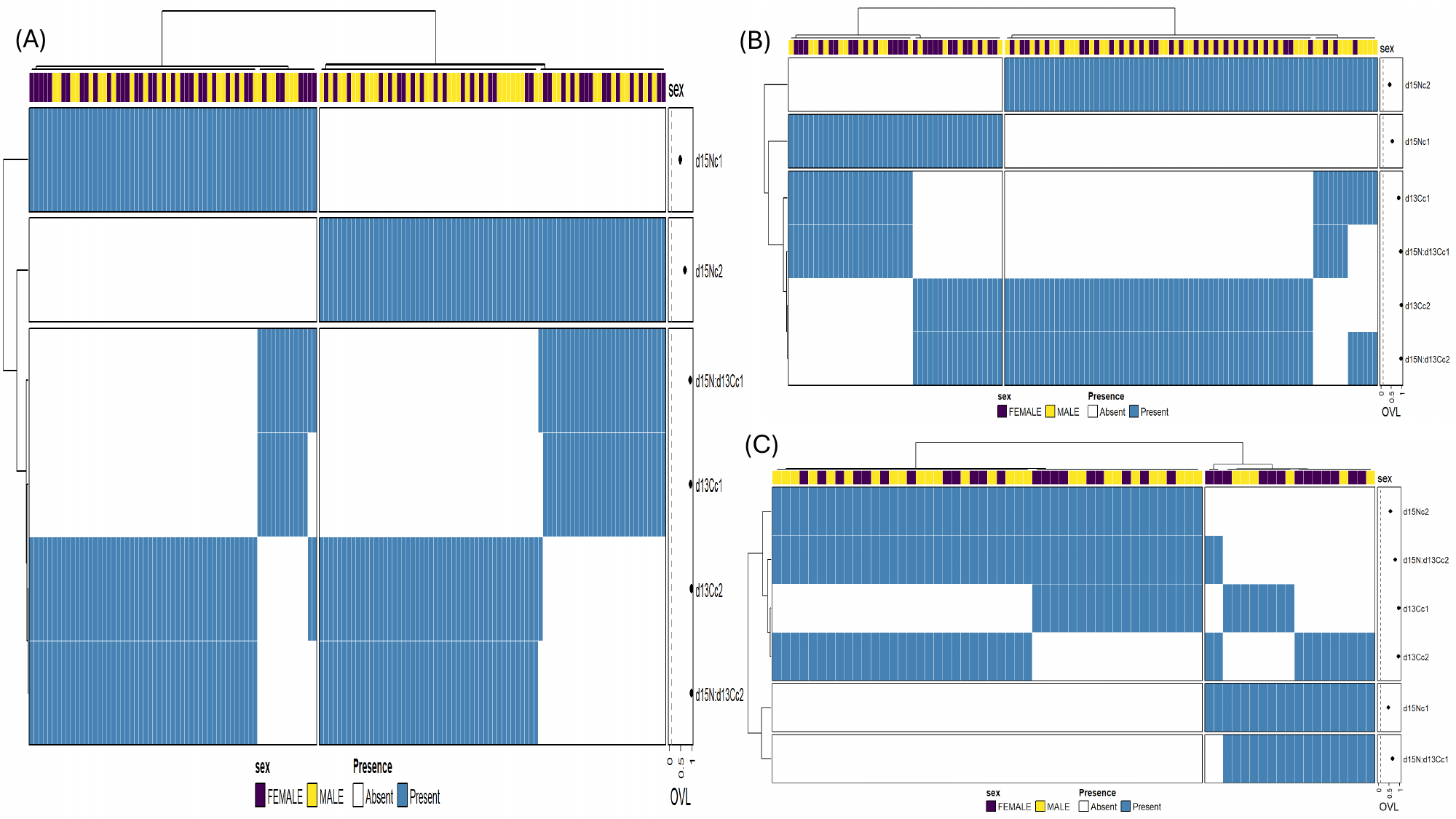}
 \caption{Three heatmaps for Stable Isotope on Sexual differences: (A) Adelie; (B) Gentoo; (C) Chinstrap.}
 \label{SIspecies}
 \end{figure}

\paragraph{Within-species Sexual differences via Stable Isotopes.}
To address within-species Sexual differences, the three species-specific Re-Co dynamics are constructed as follows. A species-specific binary response variable: $\{male, female\}$, is to be linked with two stable isotopes, $\{\delta^{15}\mathrm{N}, \delta^{13}\mathrm{C}\}$, as two covariate features. The three species-specific heatmaps are reported via three panels: (A), (B) and (C), in Fig.~\ref{SIspecies}. With respect to all overlapping areas reported along the bar-column within each heatmap, none of the associative information pieces are significant for species sexual-differences from the perspective of stable isotopes.

Our SDA-based results are not coherent with corresponding results: ``....male chinstrap and gentoo penguins were enriched
in $\delta^{15}$N relative to females...'' as reported in the original paper \cite{gorman}. The reason behind such a difference in conclusions might be due to the fact that SDA computing does not use modeling techniques, and numbers of nests of Chinstrap and Gentoo are relatively small.

\section*{Potential mate selection criteria}
In the \texttt{penguins\_lter} data set, each penguin is annotated with a nest-ID. For instance, two members of the 6th nest are encoded with: $N6A1$, $N6A2$, where ``A1'' and ``A2'' bear no information about status of Sex. Such a nest-ID would allow us to investigate which couple's bivariate-characteristics are highly dependent. No matter if such dependence is symmetric or asymmetric, the directional dependence can shed light on mate selection criteria. This is a brand-new direction of inquiry into this Genus Pygoscelis system.

\begin{figure}[h!]
 \centering
 \includegraphics[width=1.0\textwidth]{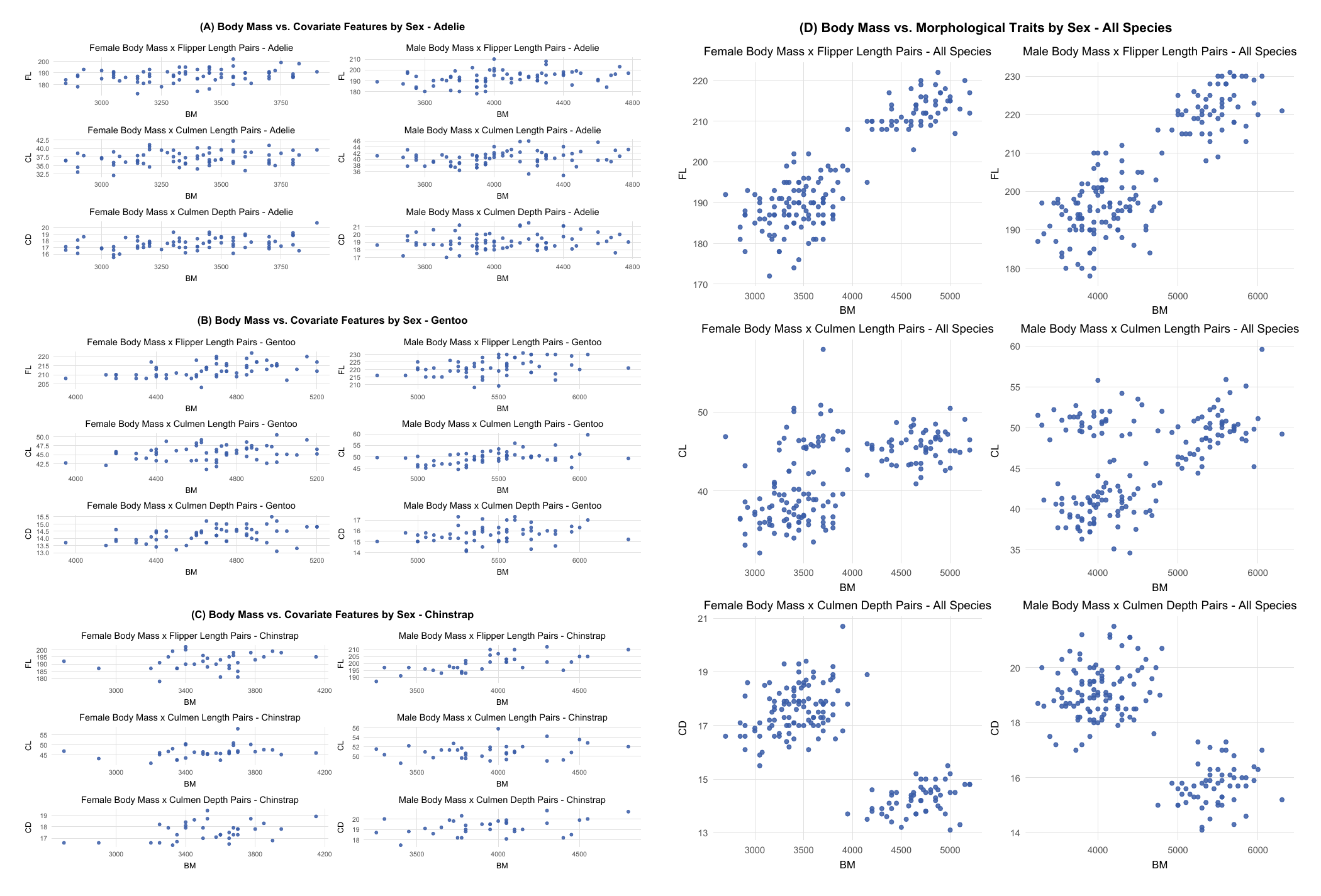}
 \caption{Males and females' BM-based 2D scatter plots: (A) Adelie; (B) Gentoo; (C) Chinstrap; (D) all penguin species.}
 \label{2Dscatterplot}
 \end{figure}

Given that we have no prior knowledge about the direction of this mate selection: either Male-choose-Female or Female-choose-Male, and whether such selection criteria are species specific or not, it is seemingly reasonable to conduct this investigation based on pooled nest data from all species. This intuitive idea needs to be validated through scatter plots of all potential covariate feature-sets of all possible orders pertaining to the set of 165 nests without missing data. We specifically illustrate scatter plots of feature-sets of order-2 belonging to the chief factor for SSD. As shown in Fig.~\ref{2Dscatterplot}, the heterogeneity becomes evident by comparing three species-specific scatter plots of $\{(BM,CD),(BM,FL),(BM,CL)\}$ in the first three panels: (A), (B) and (C) with the one based on pooled nests data across three species in the panel (D). We see explicit heterogeneity via apparently separable species-specific data-clouds pertaining to female and to male, respectively.

\begin{figure}[h!]
 \centering
 \includegraphics[width=1.0\textwidth]{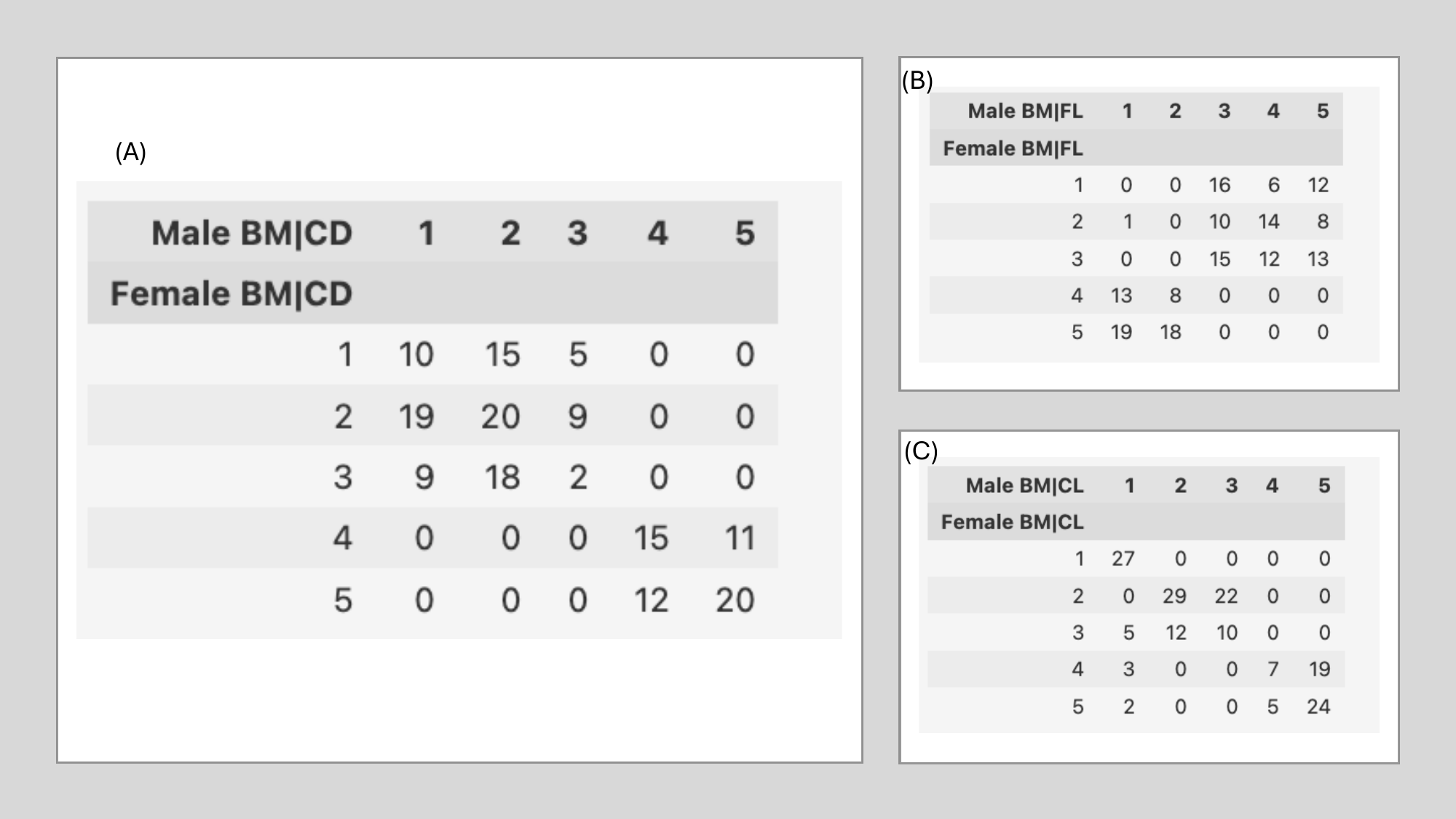}
 \caption{Order-2 contingency Tables based on nests of all penguin species: (A) (BM,CD); (B) (BM,FL); (C) (BM,CL).}
 \label{2Dcontingency}
 \end{figure}

Such evident heterogeneity characterized by species-IDs will provide biased mate-selection criteria based on the computing protocol proposed below. For example, as shown in panel (A) of Fig.~\ref{2Dcontingency}, the contingency tables of feature-set $(BM,CD)$ of order-2 based on pooled nest data, female's categories on row-axis and male's categories on column-axis are confounded with Species-IDs. That is, female's mate-selection choices are evidently tainted with Species-ID information, as are male's mate-selection choices. Similar confounding is seen in contingency tables of $(BM,FL)$ and $(BM,CL)$, respectively, in panels (B) and (C) of Fig.~\ref{2Dcontingency}. In summary, given the presence of heterogeneity visualized in Fig.~\ref{2Dscatterplot} and Fig.~\ref{2Dcontingency}, the investigation of mate-selection criteria in penguin data has to be performed within individual species.

\paragraph{Mate-selection computing protocol.}
\begin{description}
\item[a.] With respect to a covariate feature or feature-set, build two sex-specific HC-trees to categorize this covariate feature or feature-set because all covariate features are quantitative;
\item[b.] Cut both trees to have fixed number of clusters, say 3 or 5 depending on sample size, for example. Build such a contingency table with Female's categories being arranged along the row-axis, while Male's categories along the column-axis.
\item[c.] By taking the columns as 3 or 5 response categories, we confirm and collect the row-wise major feature-categories of all orders with significantly smaller entropies than the marginal one. Such major feature-categories are taken as factors of selection criteria used by Female penguins.
\item[d.] Likewise, by taking the three or five rows as response categories, we then confirm and collect the column-wise major feature-categories of all orders with significantly smaller entropies than the marginal one. Such major feature-categories are taken as factors of selection criteria used by Male penguins.
\item[e.] Collect all factors of selection criteria used by Female and arrange and display them along the row-axis of a heatmap with nest-IDs being arranged along its column-axis.
\item[f.] Collect all factors of selection criteria used by Male and arrange and display them along the row-axis of a heatmap with nest-IDs being arranged along its column-axis.
\item[g.] Compare these two heatmaps to determine ``who-choose-whom'' within this species.
\end{description}

\begin{figure}[h!]
 \centering
 \includegraphics[width=1.0\textwidth]{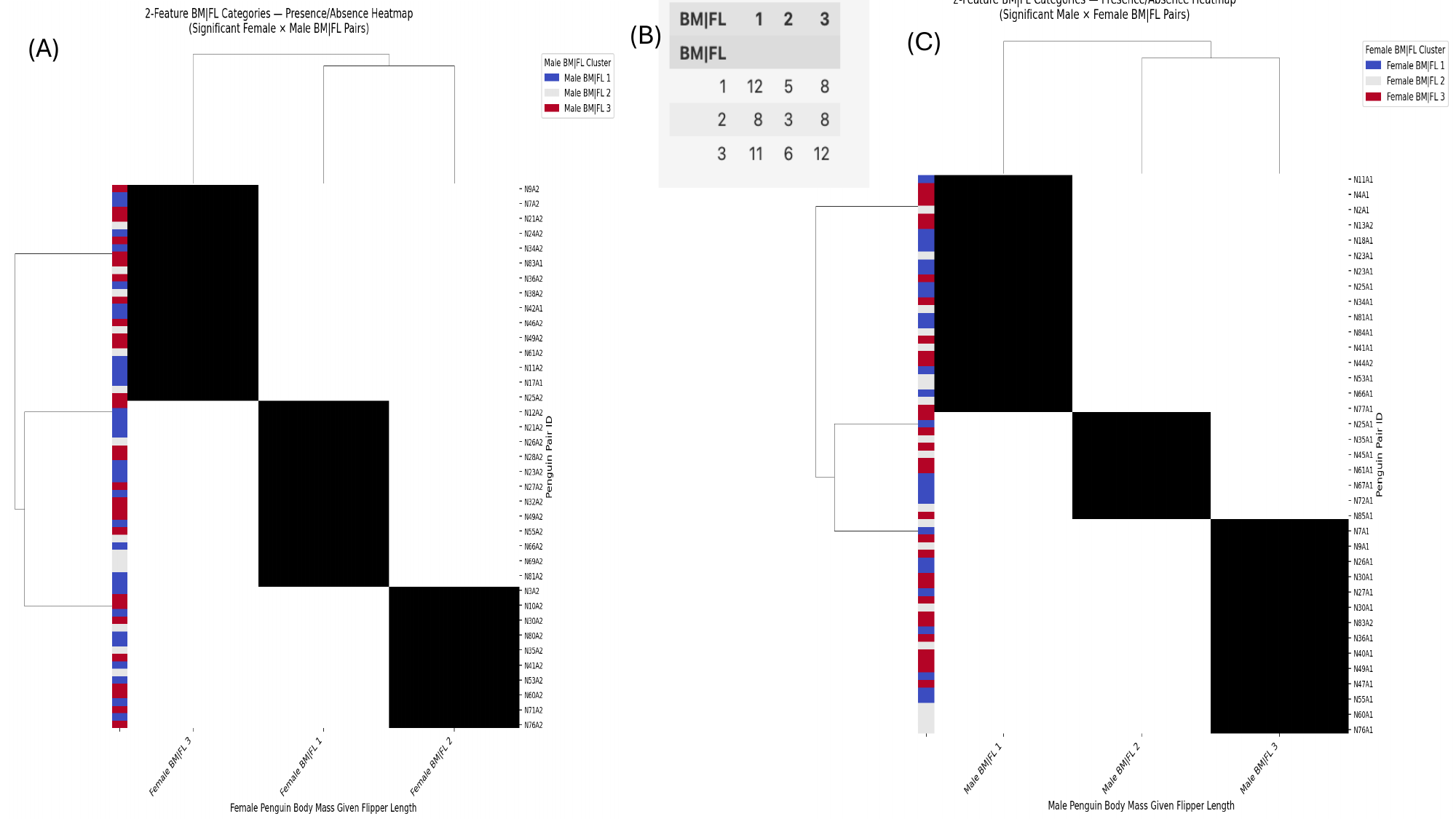}
 \caption{Order-2 $(BM, FL)$ contingency Tables of Adelie species: (A) heatmap for male-as-response-variable; (B) $3\times 3$ contingency table; (C) heatmap for female-as-response-variable.}
 \label{2DAdeliemate}
 \end{figure}

\begin{figure}[h!]
 \centering
 \includegraphics[width=1.0\textwidth]{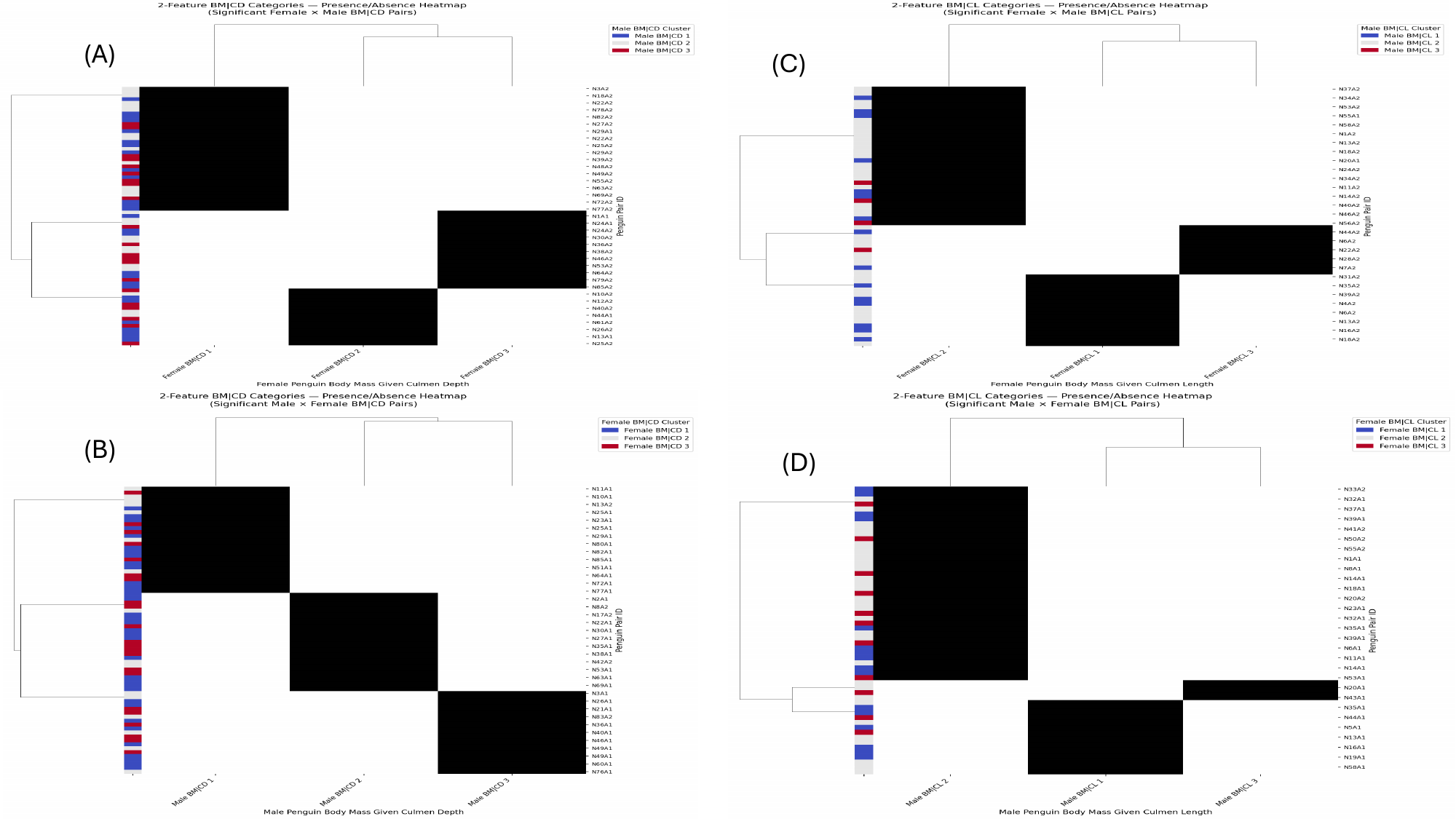}
 \caption{Order-2 $(BM, CD)$ and $(BM, CL)$ contingency Tables of Adelie species: (A) $(BM, CD)$-based heatmap for male-as-response-variable; (B) $(BM, CD)$-based heatmap for female-as-response-variable; (C) $(BM, CL)$-based heatmap for male-as-response-variable; (D) $(BM, CL)$-based heatmap for female-as-response-variable.}
 \label{2DAdeliemateBMCDCL}
 \end{figure}

We explore the potential Species-specific mate-selection criteria according to the above protocol. As shown in Fig.~\ref{2DAdeliemate}, an order-2 $(BM, FL)$ contingency table of Adelie species is given via a $3\times 3$ contingency table in the panel (B). Via SDA computations, a heatmap for male-as-response-variable is reported in panel (A), and a heatmap for female-as-response-variable in panel (C). All overlapping areas for all row-wise and column-wise associative information pieces are calculated with high values. That is, no mate-selection criteria for male and female are confirmed through $(BM, FL)$. We also confirm that no mate-selection criteria for male and female are confirmed through $(BM, CD)$ and $(BM, CL)$ through the four panels of Fig.~\ref{2DAdeliemateBMCDCL}.

\begin{figure}[h!]
 \centering
 \includegraphics[width=1.0\textwidth]{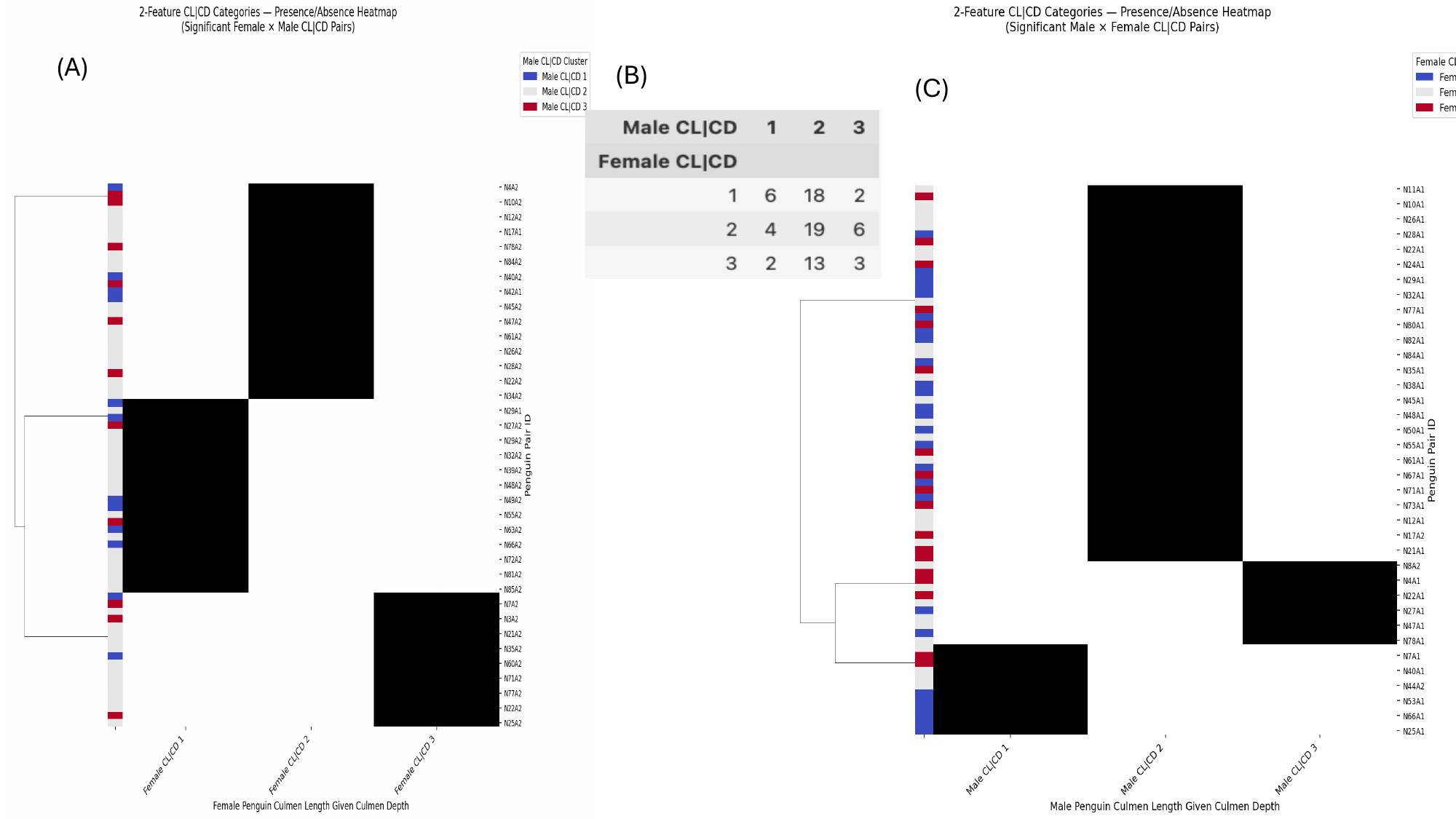}
 \caption{Order-2 $(CL, CD)$ contingency Tables of Adelie species: (A) heatmap for male-as-response-variable; (B) $3\times 3$ contingency table; (C) heatmap for female-as-response-variable.}
 \label{2DAdeliemateCDCL}
 \end{figure}

We further explore potential mate-selection criteria without involving $BM$. Our SDA computational explorations lead us to only one significant result: Adelie's male $(CL, CD)=2$, with overlapping area 0.013. The observed column vector $(18, 19, 13)$ is compared with the expected vector $(17.8, 19.86, 12.32)$ derived from the marginal counts $(26, 29, 18)$, as shown in Fig.~\ref{2DAdeliemateCDCL}. This column of male $(CL, CD)=2$ is significantly different due to the small overlapping area and large sample size.

Here, it is worth reiterating that each $3\times 3$ or $5\times 5$ contingency table presented in this section is a basis for exploring potential patterns of dependence: either from row- or column-directions. That is, through Mate-selection computing protocol, we are interested in any potential row-wise or column-wise locality-based mechanistic dependence, not overall one. That is why the common Chi-squared statistic is not suitable here. In other words, classic statistics for testing independence via Chi-squared statistic is indeed invalid in most of real-world settings from this perspective of directional dependence.

\section*{Conclusions}
We reanalyze penguin data to illustrate Computational Taxonomy (CT) as a principle of data analysis that has to be a scientific discipline as advocated by John Tukey in \cite{tukey}. The quest of CT on penguin data is to map out its heterogeneity-homogeneity with least complexity embraced within this data set. This map is multiscale in nature. As we demonstrated through the heatmap for Adelie's SSD, the male and female both embrace obvious heterogeneity. Upon this fact, Logistic regression models are not valid in modeling SSD of Adelie. Gentoo and Chinstrap species both also embrace heterogeneity of varying degrees. As such, the reported results and conclusions based on such statistical analysis in the original paper \cite{gorman} are not valid.

In sharp contrast, the answers to SSD across three penguin species bring out SSD's underlying dynamics through block-sustained heatmaps. Two major kinds of block patterns are manifested by one chief and two minor factors. The chief factor consists of 8 feature-sets from order-1 to order-4 centering around $BM$ as the dominant force, while the minor factors consist only of low-order features and feature-sets of $\{CD, CL, FL\}$. The chief factor gives rise to mechanistic dependence with very strict and exclusive block-chains of 1's and 0's blocks. In contrast, the minor factors give rise to mechanistic dependence with relatively loose, but still visible block-patterns. The explanations on SSD via heatmap-based dynamics is rather informative. The mechanistic dependence pertaining to these factors are not mentioned in \cite{gorman}. Even the dominance of $BM$ is missing.

It is worth reiterating that, through SDA applied on Penguin data set, the heterogeneity-vs-homogeneity map motivated by CT is seen truly vital for validating biological questions of SSD and mate-selection, and at the same time for deriving resultant dynamics-based answers to these two questions. Such a principle of data analysis indeed is applicable to all data analysis. Our confidence on this statement lies in the restrictive nature of the predictive modeling structure, which typically assumes homogeneity. These models fail to address the heterogeneity that naturally arises in biological systems, leading to biased results.

Conversely, CT is primarily based on SDA which is entirely free from the curse of dimensionality. Handling both quantitative vs. qualitative data becomes a non-issue, given non-categorical variables can easily be categorized via the Hierarchical Clustering (HC) algorithm with respect to their histogram boundaries \cite{FR2018}. Upon a contingency table as SDA's computing platform, each explored piece of Co-locality associative information is coupled with its idiosyncratic finite-sample precision motivated by Kolmogorov's randomness proper concept.

Hence, a Taxonomic Hierarchy with properly identified classes as its ending nodes in any CT quest will represent the targeted complex system with finite-sample legitimacy. All CT results are readable, visible and explainable. This method is applicable to any field of science.

We end this paper by explaining the opening statement of the Introduction in the original paper \cite{gorman} through CT. \\
``Intra-population variation in ecological niche, sensu \cite{hutchinson}, is widespread in nature \cite{smith,bolnick} however, the ecological and evolutionary causes and consequences of such individual variability remain poorly understood \cite{dall}.''\\
In this paper, our CT and its full version of Taxonomic hierarchy explicitly indicate that Intra-population variation is multiscale. The Intra-Sex variations: male and female, with varying complexity are particularly interesting and important phenomena. It is worth looking into in this Pygoscelis system.

In summary, our Computational Taxonomy (CT) resolves the quest of reanalyzing the penguin data set \texttt{penguins\_lter}, by providing heatmap-based SSDs with block-chains to characterize Intra-Sex variation within each species, and simultaneously offering detailed individual character-landscapes for all penguins across three species. Our final version of Taxonomic Hierarchy, which is the much expanded version from the tree with 6 $(Species, Sex)$ branches, further manifests the system-wise heterogeneity-vs-homogeneity map that represent the whole system under study. We believe that individual variability and Intra-population would be understood much better in a systematic fashion through CT.

\section*{Acknowledgements}
The authors thank the Palmer Station Long-Term Ecological Research (LTER) program and K.~B.~Gorman and colleagues for making the original penguin data set publicly available. H.~Fushing is supported by a small grant from the L\&S College of UC Davis.

\section*{Author contributions statement}
T.B.M. and H.F. conceived the study and designed the Computational Taxonomy analysis. T.B.M. and A.C.C. performed the computations and prepared the figures. T.B.M. and H.F. interpreted the results and wrote the manuscript. All authors reviewed the manuscript.

\section*{Additional information}
\noindent\textbf{Competing interests:} The authors declare no competing interests.

\end{document}